\documentclass[twocolumn]{article}

\usepackage[english]{babel}
\usepackage[top=3cm,bottom=2cm,left=3cm,right=3cm,marginparwidth=1.75cm]{geometry}
\usepackage{amsmath}
\usepackage{amsfonts}
\usepackage{graphicx}
\usepackage{tikz}
\usetikzlibrary{arrows.meta}

\newcommand{\cyc}{\raisebox{-1.5pt}{$\; \overrightarrow{\leftarrow} \;$}}

\title{FASK with Interventional Knowledge Recovers Edges from the Sachs Model}

\author{Joseph Ramsey \thanks{Carnegie Mellon University, Department of Philosophy, jdramsey@andrew.cmu.edu} \and Bryan Andrews \thanks{University of Pittsburgh, Intelligent Systems Program, BJA43@pitt.edu}}

\begin{document}
\maketitle

\begin{abstract}
\noindent We report a procedure that, in one step from continuous data with minimal preparation, recovers the graph found by Sachs et al. \cite{sachs2005causal}, with only a few edges different. The algorithm, Fast Adjacency Skewness (FASK), relies on a mixture of linear reasoning and reasoning from the skewness of variables; the Sachs data is a good candidate for this procedure since the skewness of the variables is quite pronounced. We review the ground truth model from Sachs et al. as well as some of the fluctuations seen in the protein abundances in the system, give the Sachs model and the FASK model, and perform a detailed comparison. Some variation in hyper-parameters is explored, though the main result uses values at or near the defaults learned from work modeling fMRI data.
\end{abstract}

\section{Introduction}

Sachs et al. \cite{sachs2005causal} describe a procedure that recovers most of the edges in the ground truth for a single cell biological model of protein interactions. The Sachs data consists of nine files with varying interventions (one interventional context per file). There are between 700 and 900 data points in each file, for a total of 7466 data points. These represent fluctuation in biochemical concentrations in single cells. They are all measurements of the same variables: Raf, Mek, Plc, PIP2, PIP3, Erk, Akt, Pka, Pkc, P38, Jnk. We use the names of the variables that Sachs et al. use in their article; the names in the supplied data files are in some cases different. For descriptions of these variables and the underlying biology, see \cite{sachs2005causal}. The goal of our analysis is to recover the ground truth that Sachs et al. provide (described below) and to find a model similar to the one found using their procedure.

We report a procedure that finds essentially the same model (with a few edges in difference) using the continuous version of the data and information pertaining to the interventions performed on the data. Unlike the procedure reported by Sachs et al, the method reported here find the model without discretizing, excluding points, heuristic strategies, multiple estimations, or model averaging. The algorithm, Fast Adjacency Skewness (FASK) \cite{sanchez2018causal}, is run on the Sachs data augmented with intervention variables and background knowledge which forbids edges from measured variables to intervention variables and edges from intervention variables to other intervention variables. We first explain the ground truth, then give the Sachs and FASK models showing how they relate to ground truth and each other. For a description and justification of the FASK algorithm, see \cite{sanchez2018causal}.

Although many people have analyzed this data with various methods, very few results have ended up in the public domain \cite{friedman2008sparse, aragam2017learning, henao2011sparse, miller2012identifying, desgranges2015generalization, magliacane2016joint, goudet2017causal, kalainathan2018sam}.\footnote{Models from these papers are reproduced in the Appendix.} The best of these in terms of recovering the adjacencies in the ground truth described by Sachs et al. is \cite{miller2012identifying}, but the directions of influence were not estimated. For most, the adjacencies were either very sparse or quite different from the ground truth. Where orientations are reported close to the ground truth, adjacencies have traditionally been sparse. The best result we have found in the literature, bar none, with respect to the ground truth, has been Sachs et al.’s original result \cite{sachs2005causal}. We aim to produce a result similar to theirs but with a faster and less heuristic approach.

\section{Ground Truth}

\begin{figure}
\centering
\includegraphics[width=0.48\textwidth]{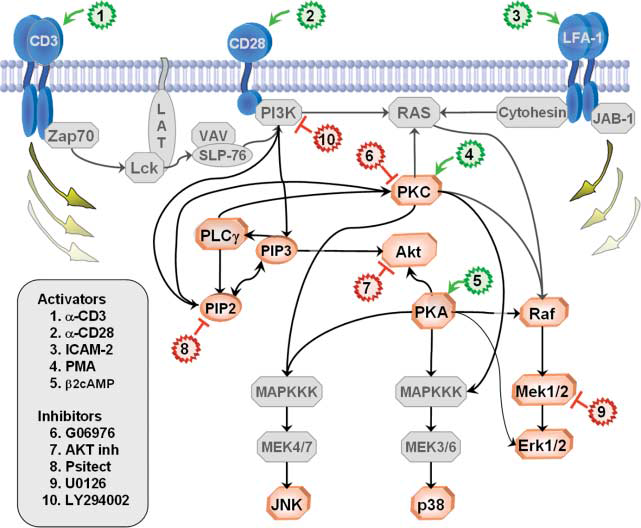}
\caption{The Biologists' View \cite{sachs2005causal}.}
\label{fig1}
\end{figure}

It is somewhat unclear as to what the ground truth is in Sachs et al \cite{sachs2005causal}. Most of the edges are not problematic but some require thought. Figure \ref{fig1} shows the Sachs et al. figure detailing the biologists’ view.

\begin{figure}
\centering
\includegraphics[width=0.48\textwidth]{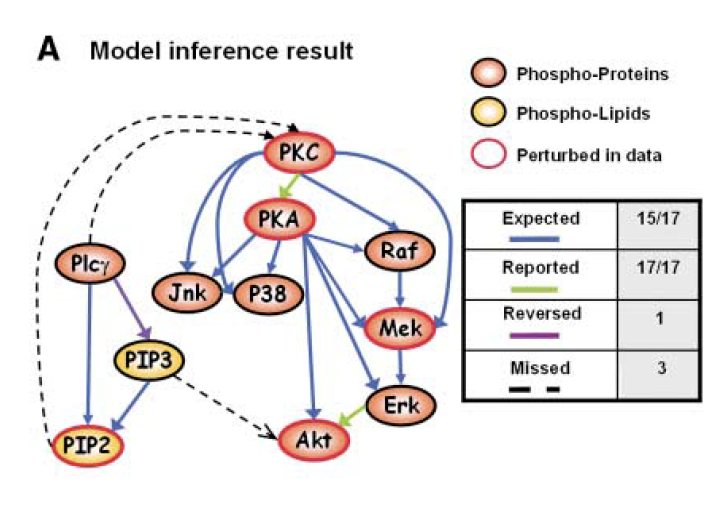}
\caption{Sachs et al.'s diagram showing a comparison of their model to their primary ground truth \cite{sachs2005causal}.}
\label{fig2}
\end{figure}

Figure \ref{fig2} shows their figure comparing their model to the ground truth.

Most of the edges in the ground truth graphs are included in the biologists’ view, above, if not as direct edges, then as indirect influences going through other (latent) variables. Figure \ref{fig2} shows Sachs et al.’s Figure 3A \cite{sachs2005causal}, giving their search result and also showing their primary ground truth. Edges in blue are edges where their model agrees with ground truth. These are edges that are reported widely throughout the literature. Dashed edges are edges in their ground truth that their model misses. Magenta edges in their model are edges that are reversed in the primary ground truth. Green edges are edges that are reported but not widely. If one starts with their Figure 3A and reverses Plc $\rightarrow$ PIP2 as indicated, that leaves the two edges in green ``reported but not expected'' and one in black. For the green edges, the authors state that perhaps we can rely on their adjacency but not their orientation. However, the authors perform an experiment confirming the orientation Erk $\rightarrow$ Akt and confirming that Erk does not cause Pka. This suggests that Erk $\rightarrow$ Akt should be included in the ground truth and that one should double check results to make sure that Erk does not cause Pka in the output graphs of search algorithms. The authors do not in their paper produce an experiment to show that Pkc causes Pka, and it is not included in the biologists’ view. This suggests that the perhaps the edge Pkc $-$ Pka should be included in the ground truth and remain unoriented. The status of the black edge Pka $\rightarrow$ Erk is less certain; the legend says that this edge is unconfirmed, and the text says as much as well. However, it is included in the biologists’ view, so there is a prima facie argument to include it in the ground truth.

\begin{figure}
\centering
\includegraphics[width=0.48\textwidth]{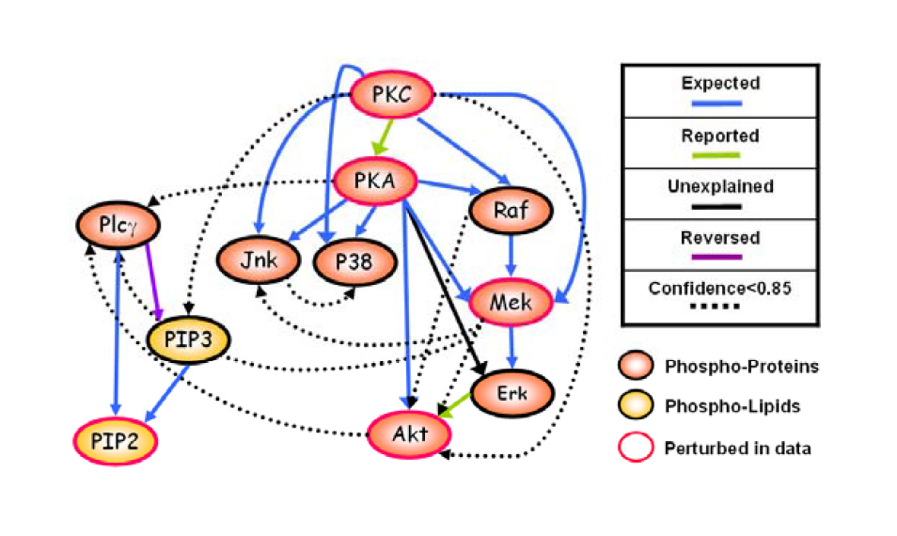}
\caption{Sachs et al.'s model compared to the supplemented ground truth \cite{sachs2005causal}.}
\label{fig3}
\end{figure}

In Sach et al.'s supplement Figure SOM3 \cite{sachs2005causal}, shown in Figure \ref{fig3}, Sachs et al. add additional ``low confidence'' edges. It seems proper to take the same attitude toward these as with the edges marked in green in Figure \ref{fig2}; perhaps the adjacencies can be asserted but not the orientations. This suggests supplementing the ground truth with the following unoriented edges to highlight correspondence with the Sachs result:

\begin{itemize}
\itemsep0em
\item Pkc -- Akt
\item Raf -- Akt
\item Mek -- Akt
\item Akt -- Plc
\item Mek -- Plc
\item Mek -- Jnk
\item Pka -- Plc
\item Jnk -- P38
\end{itemize}

These considerations lead to the graph shown in Figure \ref{fig3}, which we call the ``supplemented ground truth''.

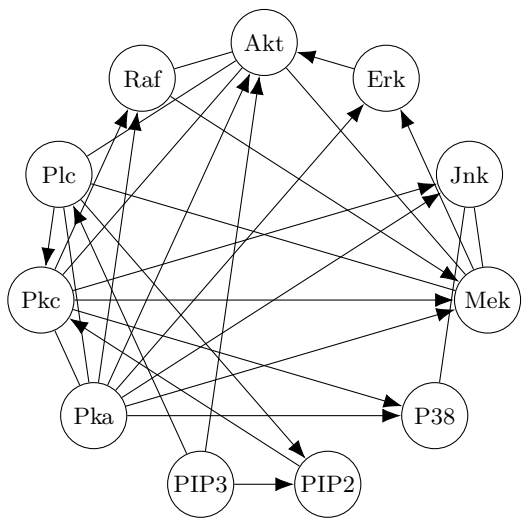
\begin{figure}
\centering
\begin{tikzpicture}
\tikzset{vertex/.style = {circle, draw, minimum size=2.5em,inner sep=1pt}}
\tikzset{edge/.style = {->, arrows={-> [scale=2]}}}
\tikzset{u_edge/.style = {->, arrows={- [scale=2]}}}
\node[vertex] (raf) at (-1.622, 2.524) {\small Raf};
\node[vertex] (mek) at (2.969, -0.427) {\small Mek};
\node[vertex] (plc) at (-2.729, 1.246) {\small Plc};
\node[vertex] (PIP2) at (0.845, -2.878) {\small PIP2};
\node[vertex] (PIP3) at (-0.845, -2.878) {\small PIP3};
\node[vertex] (erk) at (1.622, 2.524) {\small Erk};
\node[vertex] (akt) at (0.0, 3.0) {\small Akt};
\node[vertex] (pka) at (-2.267, -1.965) {\small Pka};
\node[vertex] (pkc) at (-2.969, -0.427) {\small Pkc};
\node[vertex] (p38) at (2.267, -1.965) {\small P38};
\node[vertex] (jnk) at (2.729, 1.246) {\small Jnk};
\draw[-{Latex[length=2.5mm]}] (erk) to (akt);
\draw[-{Latex[length=2.5mm]}] (mek) to (erk);
\draw[-{Latex[length=2.5mm]}] (PIP2) to (pkc);
\draw[-{Latex[length=2.5mm]}] (PIP3) to (akt);
\draw[-{Latex[length=2.5mm]}] (PIP3) to (PIP2);
\draw[-{Latex[length=2.5mm]}] (PIP3) to (plc);
\draw[-{Latex[length=2.5mm]}] (pka) to (akt);
\draw[-{Latex[length=2.5mm]}] (pka) to (erk);
\draw[-{Latex[length=2.5mm]}] (pka) to (jnk);
\draw[-{Latex[length=2.5mm]}] (pka) to (mek);
\draw[-{Latex[length=2.5mm]}] (pka) to (p38);
\draw[-{Latex[length=2.5mm]}] (pka) to (raf);
\draw[-{Latex[length=2.5mm]}] (pkc) to (jnk);
\draw[-{Latex[length=2.5mm]}] (pkc) to (mek);
\draw[-{Latex[length=2.5mm]}] (pkc) to (p38);
\draw[-] (pkc) to (pka);
\draw[-{Latex[length=2.5mm]}] (pkc) to (raf);
\draw[-{Latex[length=2.5mm]}] (plc) to (PIP2);
\draw[-{Latex[length=2.5mm]}] (plc) to (pkc);
\draw[-{Latex[length=2.5mm]}] (raf) to (mek);
\draw[-] (pkc) to (akt);
\draw[-] (raf) to (akt);
\draw[-] (mek) to (akt);
\draw[-] (akt) to (plc);
\draw[-] (mek) to (plc);
\draw[-] (mek) to (jnk);
\draw[-] (pka) to (plc);
\draw[-] (jnk) to (p38);
\end{tikzpicture}
\caption{The supplemental ground truth.}
\label{fig4}
\end{figure}

The supplemented ground truth (Figure \ref{fig4}) adds edges suggested by Sachs et al. in their Figures 3A and SOM3 \cite{sachs2005causal}. The PIP2 $-$ PIP3 edge is marked with a bidirected edge in the biologists’ view. A survey of the literature suggests that this bidirected edge is intended to represent a 2-cycle. We use the direction suggested by Sachs et al. but make comments on 2-cycles later.

\section{Sachs et al.'s Procedure}

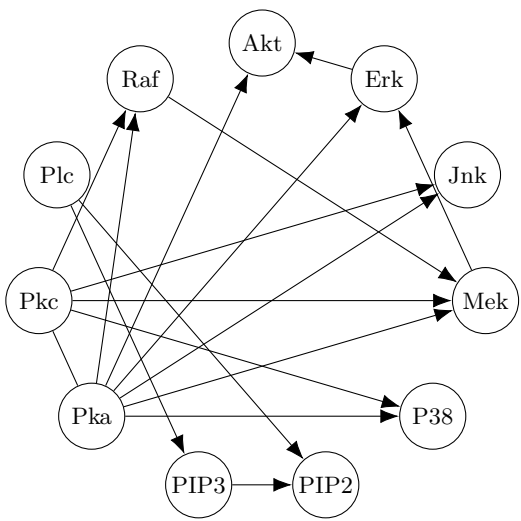
\begin{figure}
\centering
\begin{tikzpicture}
\tikzset{vertex/.style = {circle, draw, minimum size=2.5em,inner sep=1pt}}
\tikzset{edge/.style = {->, arrows={-> [scale=2]}}}
\tikzset{u_edge/.style = {->, arrows={- [scale=2]}}}
\node[vertex] (raf) at (-1.622, 2.524) {\small Raf};
\node[vertex] (mek) at (2.969, -0.427) {\small Mek};
\node[vertex] (plc) at (-2.729, 1.246) {\small Plc};
\node[vertex] (PIP2) at (0.845, -2.878) {\small PIP2};
\node[vertex] (PIP3) at (-0.845, -2.878) {\small PIP3};
\node[vertex] (erk) at (1.622, 2.524) {\small Erk};
\node[vertex] (akt) at (0.0, 3.0) {\small Akt};
\node[vertex] (pka) at (-2.267, -1.965) {\small Pka};
\node[vertex] (pkc) at (-2.969, -0.427) {\small Pkc};
\node[vertex] (p38) at (2.267, -1.965) {\small P38};
\node[vertex] (jnk) at (2.729, 1.246) {\small Jnk};
\draw[-{Latex[length=2.5mm]}] (erk) to (akt);
\draw[-{Latex[length=2.5mm]}] (mek) to (erk);
\draw[-{Latex[length=2.5mm]}] (PIP3) to (PIP2);
\draw[-{Latex[length=2.5mm]}] (pka) to (akt);
\draw[-{Latex[length=2.5mm]}] (pka) to (erk);
\draw[-{Latex[length=2.5mm]}] (pka) to (jnk);
\draw[-{Latex[length=2.5mm]}] (pka) to (mek);
\draw[-{Latex[length=2.5mm]}] (pka) to (p38);
\draw[-{Latex[length=2.5mm]}] (pka) to (raf);
\draw[-{Latex[length=2.5mm]}] (pkc) to (jnk);
\draw[-{Latex[length=2.5mm]}] (pkc) to (mek);
\draw[-{Latex[length=2.5mm]}] (pkc) to (p38);
\draw[-] (pkc) to (pka);
\draw[-{Latex[length=2.5mm]}] (pkc) to (raf);
\draw[-{Latex[length=2.5mm]}] (plc) to (PIP2);
\draw[-{Latex[length=2.5mm]}] (plc) to (PIP3);
\draw[-{Latex[length=2.5mm]}] (raf) to (mek);
\end{tikzpicture}
\caption{The Sachs et al. model \cite{sachs2005causal}.}
\label{fig5}
\end{figure}

To produce the model shown in Figure \ref{fig2}, Sachs et al. \cite{sachs2005causal} use the following procedure, detailed in the supplement of their article. They combine the nine datasets\footnote{Nagarajan et al. \cite{nagarajan2013bayesian} suggest there was a tenth dataset used to produce the $N = 5400$ discrete preparation of the data. This dataset is not available in the supplement to Sachs et al. \cite{sachs2005causal}.} in the following manner. First, they exclude points greater than three standard deviations form the mean and then discretize the data using three categories with an agglomerative method. The final discretized dataset has 5400 records, down from 7466 in the combined data from all nine datasets. They maximize a standard Bayesian score \cite{heckerman1998tutorial} by repeating a heuristic method 500 times. For each iteration, they start with a random directed acyclic graph (DAG) over the variables and then randomly add, remove, or reverse edges without violating acyclicity. At each step, if the score improves, they move to the new graph. Sometimes, to avoid a local minimum, they move to a new graph with a lower score. They perform model averaging over the 500 returned DAGs, including an edge in the final model if it occurs in at least 85\% of the 500 models. The model they end up with is shown in Figure \ref{fig5}. The procedure is detailed in Nagarajan et al., page 47 \cite{nagarajan2013bayesian}; their code examples are provided in the bnlearn package\footnote{http://www.bnlearn.com/} to help the reader approximately reproduce the original results. It is pointed out in \cite{nagarajan2013bayesian} that interventional knowledge is treated by Sachs et al. as prior information, that is, as adjustments to the Bayesian scores used to score model.  This clarifies how Sachs et al. can infer directions in places where Bayesian networks generally can only infer adjacencies.

In this report, we will use the following statistics:

\begin{itemize}
\item AP - Adjacency Precision
\item AR - Adjacency Recall
\item AHP - Arrowhead Precision; not penalizing bidirected edges
\item AHR - Arrowhead Recall
\end{itemize}

\noindent With respect to the ground truth, the Sachs model on our measures performs as follows:

\begin{center}
\begin{tabular}{ c | c | c | c }
AP & AR & AHP & AHR \\
\hline
1.00 & 0.85 & 0.94 & 0.79 
\end{tabular}
\end{center}

\noindent The 85\% for AR is consistent with Sachs et al.’s conclusions. With respect to the supplemented ground truth, the Sachs model on our measures performs as follows:

\begin{center}
\begin{tabular}{ c | c | c | c }
AP & AR & AHP & AHR \\
\hline
1.00 & 0.61 & 0.94 & 0.75
\end{tabular}
\end{center}

\section{FASK Results}

The idea of the FASK algorithm is as follows. First, FAS-stable is run on the data (this is the adjacency search of the PC-Stable algorithm, \cite{colombo2014order}), producing an undirected graph. We use the linear, Gaussian BIC score as a conditional independence test and a specified penalty discount $c$. This score assumes the data are Gaussian, but is tolerant of non-Gaussianities. This yields an undirected graph $\mathcal{G}_0$. The reason FAS-stable works for sparse cyclic models where the linear coefficients are all much less than 1 is that correlations induced by long cyclic paths are statistically judged to be zero, since they are products of multiple correlations significantly less than 1.\footnote{The adjacency step of the PC algorithm \cite{spirtes2000causation} is the same as the adjacency search for the Cyclic Causal Discovery (CCD) algorithm \cite{richardson1996discovery}; its application to cyclic models is proven there.} Then, each of the $X - Y$ adjacencies in $\mathcal{G}_0$ is oriented as a 2-cycle $X \cyc Y$, or $X \rightarrow Y$, or $X \leftarrow Y$. Taking up each adjacency in turn, FASK checks to see whether the adjacency is a 2-cycle by testing if the difference between $\text{corr}(X, Y)$ and $\text{corr}(X, Y | X > 0)$, and the differnce between $\text{corr}(X, Y)$ and $\text{corr}(X, Y | Y > 0)$, are both significantly not zero, conditioning further on all subsets of $adj(X) \backslash \{Y\}$ or $adj(Y) \backslash \{X\}$. If so, edges $X \rightarrow Y$ and $X \leftarrow Y$ are added to the output graph $\mathcal{G}_1$. If not, the Left-Right orientation rule is applied: Orient $X \rightarrow Y$ in $\mathcal{G}_1$, if 
\[
\begin{split}
&\frac{\mathbb{E}(XY | X > 0)}{ \sqrt{\mathbb{E}(X^2 | X > 0) \mathbb{E}(Y^2 | X > 0)}} \\
&- \frac{\mathbb{E}(XY | Y > 0)}{\sqrt{\mathbb{E}(X^2|Y > 0)E(Y^2 | Y > 0)}} > 0;
\end{split}
\]
otherwise orient $X \leftarrow Y$. $\mathcal{G}_1$ will be a fully oriented graph. For some models, where the true coefficients of a 2-cycle between $X$ and $Y$ are more or less equal in magnitude but opposite in sign, a correlation test may eliminate the edge between $X$ and $Y$ when in fact a 2-cycle exists. In these cases, we check explicitly whether $\text{corr}(X, Y | X > 0)$ and $\text{corr}(X, Y | Y > 0)$ differ by more than a set amount of 0.3. If so, the adjacency is added to the graph and oriented using the aforementioned rules.

\begin{figure}
\centering
\includegraphics[width=0.48\textwidth]{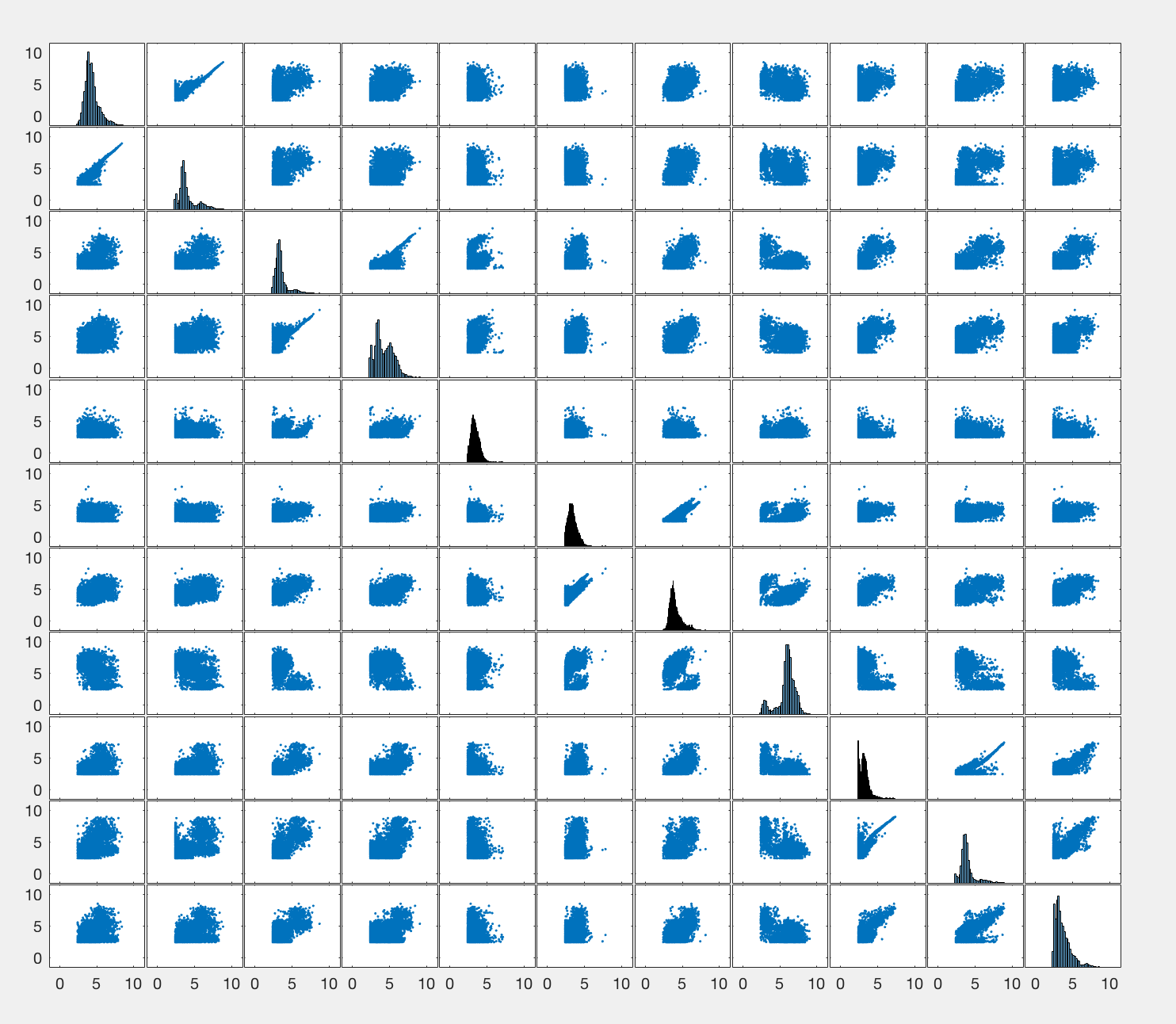}
\caption{Scatter plots for the logged, transformed data. This is the order of the variables: Raf, Mek, Plc, PIP2, PIP3, Erk, Akt, Pka, Pkc, P38, Jnk.}
\label{fig6}
\end{figure}

For our analysis, we use all 7466 records of the data, prepared as follows. Since interventional data is available, we add this to the data. We create a column in the data for each specific compound introduced (as noted in the names of the ``main result'' files in \cite{sachs2005causal}) and put a 1 in rows where that compound is introduced 0 otherwise.  We merge the columns for the chemicals cd3 and cd28 since they always co-occur. 
We then jiggle the intervention data by adding draws from a normal distribution $N(0, 0.01)$, to avoid singularities. We take the view (as Sachs et al. \cite{sachs2005causal} do) that the protein data may be concatenated, if intervention data is taken into account. The reason for this is that all protein level measurements are carried out using the same procedure and thus are comparable. We log the data (after adding 10, so applying the transformation $f(x) = \log(10 + x)$) and generate a concatenated dataset including all of the jiggled intervention data and logged data.\footnote{This dataset is attached as ``data.txt''.} We then apply the FASK algorithm to this dataset. The data supplied by Sachs et al. \cite{sachs2005causal} are very skewed, making them a good candidate for this procedure. The log transformation applied to the data makes them somewhat more Gaussian for the adjacency search, while leaving them sufficiently skewed for the orientation search. Figure \ref{fig6} shows pairwise scatter plots over the measured variables for the resulting combined dataset.

\begin{figure}
\centering
\begin{tikzpicture}
\tikzset{vertex/.style = {circle, draw, minimum size=2.5em,inner sep=1pt}}
\tikzset{edge/.style = {->, arrows={-> [scale=2]}}}
\tikzset{u_edge/.style = {->, arrows={- [scale=2]}}}
\node[vertex] (raf) at (-1.622, 2.524) {\small Raf};
\node[vertex] (mek) at (2.969, -0.427) {\small Mek};
\node[vertex] (plc) at (-2.729, 1.246) {\small Plc};
\node[vertex] (PIP2) at (0.845, -2.878) {\small PIP2};
\node[vertex] (PIP3) at (-0.845, -2.878) {\small PIP3};
\node[vertex] (erk) at (1.622, 2.524) {\small Erk};
\node[vertex] (akt) at (0.0, 3.0) {\small Akt};
\node[vertex] (pka) at (-2.267, -1.965) {\small Pka};
\node[vertex] (pkc) at (-2.969, -0.427) {\small Pkc};
\node[vertex] (p38) at (2.267, -1.965) {\small P38};
\node[vertex] (jnk) at (2.729, 1.246) {\small Jnk};
\draw[-{Latex[length=2.5mm]}, bend left=12] (akt) to (pka);
\draw[-{Latex[length=2.5mm]}] (erk) to (akt);
\draw[-{Latex[length=2.5mm]}] (jnk) to (mek);
\draw[-{Latex[length=2.5mm]}, bend left=12] (mek) to (raf);
\draw[-{Latex[length=2.5mm]}] (PIP3) to (PIP2);
\draw[-{Latex[length=2.5mm]}] (PIP3) to (plc);
\draw[-{Latex[length=2.5mm]}] (pka) to (akt);
\draw[-{Latex[length=2.5mm]}] (pka) to (erk);
\draw[-{Latex[length=2.5mm]}] (pka) to (jnk);
\draw[-{Latex[length=2.5mm]}] (pka) to (mek);
\draw[-{Latex[length=2.5mm]}] (pka) to (p38);
\draw[-{Latex[length=2.5mm]}] (pka) to (PIP2);
\draw[-{Latex[length=2.5mm]}] (pka) to (pkc);
\draw[-{Latex[length=2.5mm]}] (pka) to (plc);
\draw[-{Latex[length=2.5mm]}] (pka) to (raf);
\draw[-{Latex[length=2.5mm]}] (pkc) to (jnk);
\draw[-{Latex[length=2.5mm]}] (pkc) to (mek);
\draw[-{Latex[length=2.5mm]}] (pkc) to (p38);
\draw[-{Latex[length=2.5mm]}] (pkc) to (raf);
\draw[-{Latex[length=2.5mm]}] (plc) to (PIP2);
\draw[-{Latex[length=2.5mm]}] (raf) to (mek);
\end{tikzpicture}
\caption{The FASK model, using a 2-cycle alpha of $10^{-5}$.}
\label{fig7}
\end{figure}
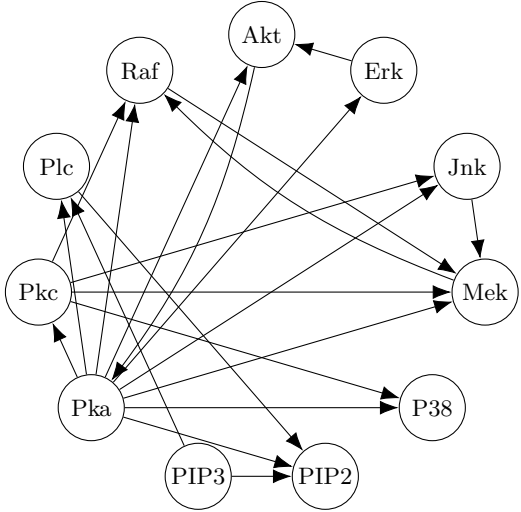

Using the prepared data as above, we created knowledge that the intervention variables are exogenous, a standard technique \cite{zhang2015discovery, magliacane2016joint}. This knowledge was supplied to the FASK algorithm along with the data augmented by intervention variables. The penalty discount for FAS was set to 1 (standard BIC). The depth of the FAS search was unlimited. After running FASK, we deleted the intervention variables from the resulting graph keeping only the graph over the measured variables. We compared this to the extended ground truth; this FASK result using all of the default parameters is as shown in Figure \ref{fig7}. 

As for choices of FASK's hyper-parameter settings, the extra edge threshold of 0.3 was a value learned from work with fMRI; this value needs to be noticeably different from zero. If it is lowered, the graph becomes more dense. The delta parameter is in the range -1 to 0; we use the default -0.2 here, though the output of FASK for this data is insensitive to particular choice of this parameter value. The 2-cycle alpha level is chosen to be a low value here, $10^{-5}$. The number of 2-cycles in the output graph varies with this value. Below we will give a result with more 2-cycles where we chose alpha = 0.05, a much higher value.

With respect to the ground truth, the FASK model performs as follows:

\begin{center}
\begin{tabular}{ c | c | c | c }
AP & AR & AHP & AHR \\
\hline
0.84 & 0.80 & 1.00 & 0.79
\end{tabular}
\end{center}

With respect to the supplemented ground truth, the FASK model performs as follows:

\begin{center}
\begin{tabular}{ c | c | c | c }
AP & AR & AHP & AHR \\
\hline
.95 & 0.64 & 1.00 & 0.79
\end{tabular}
\end{center}

Aside from rendering some adjacencies as 2-cycles, this is how the FASK model modifies the Sachs model (target graph from FASK model, True graph from Sachs model): \\

Edges added:
\begin{enumerate}
\itemsep0em
\item Pka $\rightarrow$ PIP2
\item Pka $\rightarrow$ Plc
\item Jnk $\rightarrow$ Mek
\end{enumerate}

Edges removed:
\begin{enumerate}
\itemsep0em
\item Mek $\rightarrow$ Erk
\end{enumerate}

Edges reoriented:
\begin{enumerate}
\itemsep0em
\item Pkc $-$ Pka $\Rightarrow$ Pka $\rightarrow$ Pkc
\item Plc $\rightarrow$ PIP3 $\Rightarrow$ PIP3 $\rightarrow$ Plc
\end{enumerate}

\noindent Three adjacencies are added, two of which correspond to adjacencies in the supplemented ground truth. The Pkc $\rightarrow$ Pka edge from the Sachs model is reversed, though the direction of this edge is reported but not expected. Two adjacencies are oriented as 2-cycles. A backward edge Mek $\rightarrow$ Raf edge added, rendering a 2-cycle, consistently with the literature we have surveyed \cite{leicht2013Mek}. Pka $\cyc$ Akt has the canonical direction Pka $\rightarrow$ Akt; here, it is rendered as a 2-cycle. There is at least one paper exploring cyclic models with this data \cite{itani2010structure} finds Akt $\rightarrow$ Pka.

Lastly, the Mek $\rightarrow$ Erk edge is removed. The latter is significant, since, as Sachs et al. point out, the inclusion of that edge in their model is evidence that their model has merit. Specifically, they include a path Raf $\rightarrow$ Mek $\rightarrow$ Erk; this is a well-known path in the literature. This requires a comment, which we will give below. 

\section{Comments on the FASK Model}

The Mek $\rightarrow$ Erk edge aside, the remarkable thing about Figure \ref{fig7} is how much it recovers and the degree of accuracy with which this information is recovered. No expected orientation is reversed in the model; even Sachs et al.'s \cite{sachs2005causal} analysis reverses one expected orientation. Also, although it is not expected that this will be the case for every analysis, in this case the hyper-parameter values used FASK algorithm were used to recover this information essentially the same as the defaults learned from experience with fMRI data. It is also noteworthy that Figure \ref{fig7} was obtained in one pass through the algorithm, from the logged continuous data, not a discretization of this data.

As noted, there are two types of adjacencies in FASK. First, there are adjacencies due to the Fast Adjacency Search (FAS), that is, the PC adjacency search. We use the ``-Stable'' version of this algorithm \cite{colombo2014order}. These assume the variables are distributed as approximately Gaussian. Second, there are adjacencies due to differences in correlation conditional on one or the other variable being greater than zero; the default cutoff for this difference is 0.3. As explained in Appendix A of \cite{sanchez2018causal}, these respond to skewnesses of variables in the data. We give the two subgraphs separately. Figure \ref{fig8} shows the subgraph of FAS adjacencies, oriented using the FASK orientation rules. Figure \ref{fig9} shows the subgraph of conditional correlation differences adjacencies, again oriented by FASK rules. In this case, two subgraphs are entirely disjoint and together form a graph that approximates the Sachs graph. The fact that these two sets of adjacencies are different and even disjoint suggests that an analysis of the Sachs data using algorithms that rely on the linear, Gaussian BIC score alone, or even really any test or score that relies on properties of Gaussianity alone, will not be able to recover all of the edges in the Sachs model. A survey of such algorithms verifies that this is the case. There are some edges (Figure \ref{fig9}) that can be recovered only by a method sensitive to non-Gaussianities and non-linearities in the data. In fact, in the literature we see that methods that consider non-Gaussianity and non-linearity are more successful \cite{miller2012identifying, desgranges2015generalization}.
	
The adjacency rule used in Figure \ref{fig9}, adding an adjacency between $X$ and $Y$ if
\[
| \text{corr}(X, Y | X > 0) - \text{corr}(X, Y | Y > 0) | > 0.3, 
\]
was a heuristic developed for search over fMRI data to handle the case where a control 2-cycle between $X$ and $Y$ (a cycle in which coefficients in opposite directions have opposite sign) could not be discerned from correlation because the influences in either direction canceled each other out. The idea was that this difference would be zero if $X$ and $Y$ were independent; otherwise, a trek would need to exist between $X$ and $Y$. If the difference were very different from zero (such as if the absolute difference were greater than 0.3) we heuristically infer an adjacency between $X$ and $Y$. It was a pleasant surprise to discover that this same reasoning could be put to use to find connections in other cases where linear, Gaussian methods might fail to find a connection. As a rule of thumb, for the Sachs data, this reasoning allows easy recovery of many of the ground truth edges, identified from experiment.

FASK adds three edges to the Sachs model. Two of the edges are adjacencies in the extended ground truth but not in the Sachs model. The third, Pka $\rightarrow$ PIP2 directed out of Pka; perhaps this is not surprising, given that it transitively closes a short path from Pka through Plc to PIP2, but it is a false positive adjacency with respect to the extended ground truth, so we mark it as such. FASK orients the Plc $\rightarrow$ PIP3 edge correctly; the Sachs model reverses this edge. 

Both the FASK procedure and Sachs et al.'s procedure get an adjacency between Pka and Pkc, oriented albeit in different directions. We do not feel that we have sufficient information to warrant a judgment in either direction; in fact, there is literature now claiming that cross-talk exists between Pka and Pkc, suggesting a 2-cycle. In any case, the safe thing to do for ground truth is to leave the edge unoriented, as we do. The edge Mek $-$ Jnk is oriented differently from the extended ground truth in Sachs et al. \cite{sachs2005causal}, though again, judging from the literature, we do not feel we have sufficient information to know that this is not confounded by additional variables other than the common ancestors already in the graph, so we leave this edge unoriented in the ground truth, as we do for the additional unoriented edges shown in Figure \ref{fig4}. Nevertheless, if these two edges were oriented as in Figure \ref{fig2}, the performance of the FASK model is as follows.

\begin{center}
\begin{tabular}{ c | c | c | c }
AP & AR & AHP & AHR \\
\hline
0.95 & 0.73 & 0.89 & 0.89
\end{tabular}
\end{center}

\begin{figure}
\centering
\begin{tikzpicture}
\tikzset{vertex/.style = {circle, draw, minimum size=2.5em,inner sep=1pt}}
\tikzset{edge/.style = {->, arrows={-> [scale=2]}}}
\tikzset{u_edge/.style = {->, arrows={- [scale=2]}}}
\node[vertex] (raf) at (-1.622, 2.524) {\small Raf};
\node[vertex] (mek) at (2.969, -0.427) {\small Mek};
\node[vertex] (plc) at (-2.729, 1.246) {\small Plc};
\node[vertex] (PIP2) at (0.845, -2.878) {\small PIP2};
\node[vertex] (PIP3) at (-0.845, -2.878) {\small PIP3};
\node[vertex] (erk) at (1.622, 2.524) {\small Erk};
\node[vertex] (akt) at (0.0, 3.0) {\small Akt};
\node[vertex] (pka) at (-2.267, -1.965) {\small Pka};
\node[vertex] (pkc) at (-2.969, -0.427) {\small Pkc};
\node[vertex] (p38) at (2.267, -1.965) {\small P38};
\node[vertex] (jnk) at (2.729, 1.246) {\small Jnk};
\draw[-{Latex[length=2.5mm]}] (erk) to (akt);
\draw[-{Latex[length=2.5mm]}, bend left=12] (mek) to (raf);
\draw[-{Latex[length=2.5mm]}] (PIP3) to (PIP2);
\draw[-{Latex[length=2.5mm]}] (pkc) to (jnk);
\draw[-{Latex[length=2.5mm]}] (pkc) to (p38);
\draw[-{Latex[length=2.5mm]}] (plc) to (PIP2);
\draw[-{Latex[length=2.5mm]}] (raf) to (mek);
\end{tikzpicture}
\caption{Figure 8. The subgraph of the FASK result of adjacencies due to the Fast Adjacency Search (FAS). For this, FAS was run on the variables, along with the jittered intervention variables, and FASK orientation was applied, then the jittered intervention variables were removed.}
\label{fig8}
\end{figure}
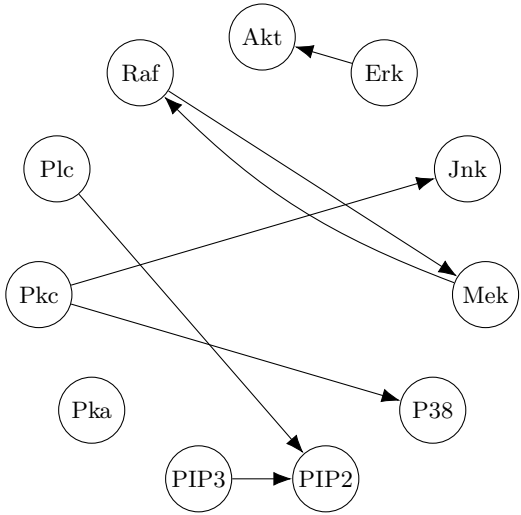

\begin{figure}
\centering
\begin{tikzpicture}
\tikzset{vertex/.style = {circle, draw, minimum size=2.5em,inner sep=1pt}}
\tikzset{edge/.style = {->, arrows={-> [scale=2]}}}
\tikzset{u_edge/.style = {->, arrows={- [scale=2]}}}
\node[vertex] (raf) at (-1.622, 2.524) {\small Raf};
\node[vertex] (mek) at (2.969, -0.427) {\small Mek};
\node[vertex] (plc) at (-2.729, 1.246) {\small Plc};
\node[vertex] (PIP2) at (0.845, -2.878) {\small PIP2};
\node[vertex] (PIP3) at (-0.845, -2.878) {\small PIP3};
\node[vertex] (erk) at (1.622, 2.524) {\small Erk};
\node[vertex] (akt) at (0.0, 3.0) {\small Akt};
\node[vertex] (pka) at (-2.267, -1.965) {\small Pka};
\node[vertex] (pkc) at (-2.969, -0.427) {\small Pkc};
\node[vertex] (p38) at (2.267, -1.965) {\small P38};
\node[vertex] (jnk) at (2.729, 1.246) {\small Jnk};
\draw[-{Latex[length=2.5mm]}, bend left=12] (akt) to (pka);
\draw[-{Latex[length=2.5mm]}] (jnk) to (mek);
\draw[-{Latex[length=2.5mm]}] (PIP3) to (plc);
\draw[-{Latex[length=2.5mm]}] (pka) to (akt);
\draw[-{Latex[length=2.5mm]}] (pka) to (erk);
\draw[-{Latex[length=2.5mm]}] (pka) to (jnk);
\draw[-{Latex[length=2.5mm]}] (pka) to (mek);
\draw[-{Latex[length=2.5mm]}] (pka) to (p38);
\draw[-{Latex[length=2.5mm]}] (pka) to (PIP2);
\draw[-{Latex[length=2.5mm]}] (pka) to (pkc);
\draw[-{Latex[length=2.5mm]}] (pka) to (plc);
\draw[-{Latex[length=2.5mm]}] (pka) to (raf);
\draw[-{Latex[length=2.5mm]}] (pkc) to (mek);
\draw[-{Latex[length=2.5mm]}] (pkc) to (raf);
\end{tikzpicture}
\caption{The subgraph of the FASK result of adjacencies due to the difference of conditional correlation difference rule. For this, the FASK heuristic adjacency rule was used to identify adjacencies among the protein level variables, and then FASK orientation was then applied.}
\label{fig9}
\end{figure}
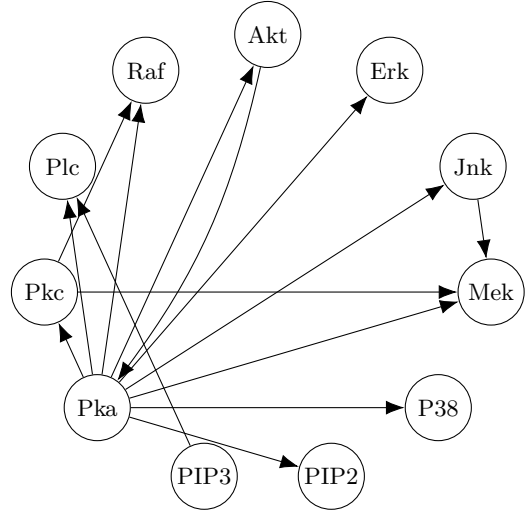

\begin{figure}
\centering
\begin{tikzpicture}
\tikzset{vertex/.style = {circle, draw, minimum size=2.5em,inner sep=1pt}}
\tikzset{edge/.style = {->, arrows={-> [scale=2]}}}
\tikzset{u_edge/.style = {->, arrows={- [scale=2]}}}
\node[vertex] (raf) at (-1.622, 2.524) {\small Raf};
\node[vertex] (mek) at (2.969, -0.427) {\small Mek};
\node[vertex] (plc) at (-2.729, 1.246) {\small Plc};
\node[vertex] (PIP2) at (0.845, -2.878) {\small PIP2};
\node[vertex] (PIP3) at (-0.845, -2.878) {\small PIP3};
\node[vertex] (erk) at (1.622, 2.524) {\small Erk};
\node[vertex] (akt) at (0.0, 3.0) {\small Akt};
\node[vertex] (pka) at (-2.267, -1.965) {\small Pka};
\node[vertex] (pkc) at (-2.969, -0.427) {\small Pkc};
\node[vertex] (p38) at (2.267, -1.965) {\small P38};
\node[vertex] (jnk) at (2.729, 1.246) {\small Jnk};
\draw[-{Latex[length=2.5mm]}, bend left=12] (akt) to (pka);
\draw[-{Latex[length=2.5mm]}] (erk) to (akt);
\draw[-{Latex[length=2.5mm]}] (jnk) to (mek);
\draw[-{Latex[length=2.5mm]}] (jnk) to (p38);
\draw[-{Latex[length=2.5mm]}] (jnk) to (PIP3);
\draw[-{Latex[length=2.5mm]}] (mek) to (akt);
\draw[-{Latex[length=2.5mm]}, bend left=12] (mek) to (raf);
\draw[-{Latex[length=2.5mm]}] (p38) to (akt);
\draw[-{Latex[length=2.5mm]}] (p38) to (pkc);
\draw[-{Latex[length=2.5mm]}] (PIP3) to (PIP2);
\draw[-{Latex[length=2.5mm]}] (PIP3) to (plc);
\draw[-{Latex[length=2.5mm]}] (pka) to (akt);
\draw[-{Latex[length=2.5mm]}] (pka) to (erk);
\draw[-{Latex[length=2.5mm]}] (pka) to (jnk);
\draw[-{Latex[length=2.5mm]}] (pka) to (mek);
\draw[-{Latex[length=2.5mm]}] (pka) to (p38);
\draw[-{Latex[length=2.5mm]}] (pka) to (PIP2);
\draw[-{Latex[length=2.5mm]}] (pka) to (pkc);
\draw[-{Latex[length=2.5mm]}] (pka) to (plc);
\draw[-{Latex[length=2.5mm]}] (pka) to (raf);
\draw[-{Latex[length=2.5mm]}] (pkc) to (jnk);
\draw[-{Latex[length=2.5mm]}] (pkc) to (mek);
\draw[-{Latex[length=2.5mm]}, bend left=12] (pkc) to (p38);
\draw[-{Latex[length=2.5mm]}] (pkc) to (raf);
\draw[-{Latex[length=2.5mm]}] (plc) to (akt);
\draw[-{Latex[length=2.5mm]}] (plc) to (jnk);
\draw[-{Latex[length=2.5mm]}] (plc) to (p38);
\draw[-{Latex[length=2.5mm]}] (plc) to (PIP2);
\draw[-{Latex[length=2.5mm]}, bend left=12] (plc) to (pka);
\draw[-{Latex[length=2.5mm]}] (raf) to (mek);
\end{tikzpicture}
\caption{The graph over FAS edges that would result if interventional knowledge were not taken into account. Orientations are inferred from skewness. This would replace Figure \ref{fig7} and contains a number of errors with respect to the extended ground truth.}
\label{fig10}
\end{figure}
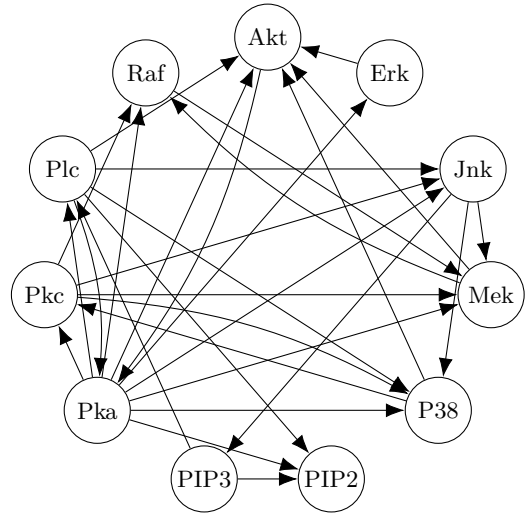

If we leave interventional knowledge out of the FASK search, we obtain the graph shown in Figure \ref{fig10}, which modifies the Sachs model as follows (target graph from FASK without interventions, true graph from Sachs model): \\

Edges added:
\begin{enumerate}
\itemsep0em
\item Plc $\rightarrow$ Pka
\item Pka $\rightarrow$ Plc
\item Plc $\rightarrow$ Pka
\item Pka $\rightarrow$ Plc
\item P38 $\rightarrow$ Akt
\item Jnk $\rightarrow$ PIP3
\item Jnk $\rightarrow$ P38
\item Plc $\rightarrow$ Akt
\item Pka $\rightarrow$ PIP2
\item Plc $\rightarrow$ P38
\item Jnk $\rightarrow$ Mek
\item Plc $\rightarrow$ Jnk
\item Mek $\rightarrow$ Akt
\end{enumerate}

Edges removed:
\begin{enumerate}
\itemsep0em
\item Mek $\rightarrow$ Erk
\end{enumerate}

Edges reoriented:
\begin{enumerate}
\itemsep0em
\item Pkc $-$ Pka $\Rightarrow$ Pka $\rightarrow$ Pkc
\item Plc $\rightarrow$ PIP3 $\Rightarrow$ PIP3 $\rightarrow$ Plc
\end{enumerate}

\noindent Essentially, this manipulation adds several edges to the graph that otherwise would not have been there. 

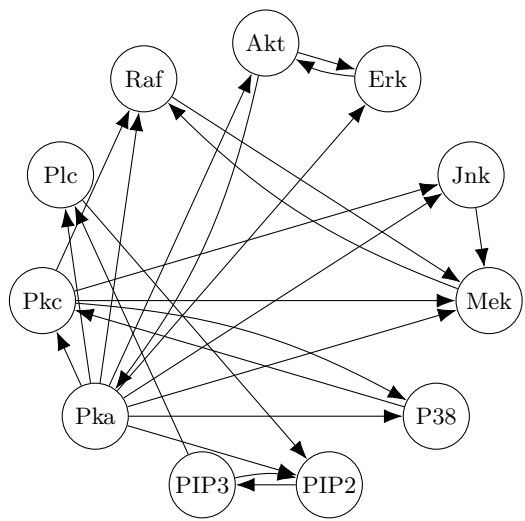
\begin{figure}
\centering
\begin{tikzpicture}
\tikzset{vertex/.style = {circle, draw, minimum size=2.5em,inner sep=1pt}}
\tikzset{edge/.style = {->, arrows={-> [scale=2]}}}
\tikzset{u_edge/.style = {->, arrows={- [scale=2]}}}
\node[vertex] (raf) at (-1.622, 2.524) {\small Raf};
\node[vertex] (mek) at (2.969, -0.427) {\small Mek};
\node[vertex] (plc) at (-2.729, 1.246) {\small Plc};
\node[vertex] (PIP2) at (0.845, -2.878) {\small PIP2};
\node[vertex] (PIP3) at (-0.845, -2.878) {\small PIP3};
\node[vertex] (erk) at (1.622, 2.524) {\small Erk};
\node[vertex] (akt) at (0.0, 3.0) {\small Akt};
\node[vertex] (pka) at (-2.267, -1.965) {\small Pka};
\node[vertex] (pkc) at (-2.969, -0.427) {\small Pkc};
\node[vertex] (p38) at (2.267, -1.965) {\small P38};
\node[vertex] (jnk) at (2.729, 1.246) {\small Jnk};
\draw[-{Latex[length=2.5mm]}] (akt) to (erk);
\draw[-{Latex[length=2.5mm]}, bend left=12] (akt) to (pka);
\draw[-{Latex[length=2.5mm]}, bend left=12] (erk) to (akt);
\draw[-{Latex[length=2.5mm]}] (jnk) to (mek);
\draw[-{Latex[length=2.5mm]}, bend left=12] (mek) to (raf);
\draw[-{Latex[length=2.5mm]}] (p38) to (pkc);
\draw[-{Latex[length=2.5mm]}] (PIP2) to (PIP3);
\draw[-{Latex[length=2.5mm]}, bend left=12] (PIP3) to (PIP2);
\draw[-{Latex[length=2.5mm]}] (PIP3) to (plc);
\draw[-{Latex[length=2.5mm]}] (pka) to (akt);
\draw[-{Latex[length=2.5mm]}] (pka) to (erk);
\draw[-{Latex[length=2.5mm]}] (pka) to (jnk);
\draw[-{Latex[length=2.5mm]}] (pka) to (mek);
\draw[-{Latex[length=2.5mm]}] (pka) to (p38);
\draw[-{Latex[length=2.5mm]}] (pka) to (PIP2);
\draw[-{Latex[length=2.5mm]}] (pka) to (pkc);
\draw[-{Latex[length=2.5mm]}] (pka) to (plc);
\draw[-{Latex[length=2.5mm]}] (pka) to (raf);
\draw[-{Latex[length=2.5mm]}] (pkc) to (jnk);
\draw[-{Latex[length=2.5mm]}] (pkc) to (mek);
\draw[-{Latex[length=2.5mm]}, bend left=12] (pkc) to (p38);
\draw[-{Latex[length=2.5mm]}] (pkc) to (raf);
\draw[-{Latex[length=2.5mm]}] (plc) to (PIP2);
\draw[-{Latex[length=2.5mm]}] (raf) to (mek);
\end{tikzpicture}
\caption{FASK model using a 2-cycle alpha of 0.05, showing five 2-cycles, some of which may be due to confounding. See text.}
\label{fig11}
\end{figure}

If, on the other hand, we leave the interventional knowledge in place and raise the 2-cycle alpha to 0.05 in the FASK search, we obtain the graph in Figure \ref{fig11}, with some additional 2-cycles. There is some rationale for each the additional backward edges. The Raf $\cyc$ Mek 2-cycle has already been discussed, as has the Pka $\cyc$ Akt 2-cycle. The PIP2 $\cyc$ PIP3 2-cycle is in fact in Figure 2A from Sachs et al. \cite{sachs2005causal}, the ``biologists’ view'' as a bidirected edge. A survey of the literature indicates that this is meant to be a 2-cycle. Pkc $\cyc$ P38 is a 2-cycle that is often found in cyclic search with this data; the cyclic paper above finds it as well. Akt $\cyc$ Erk is a little harder to explain. This is one of Sachs et al.’s ``green edges''; they perturb Erk and show that Akt responds. They do not do the opposite experiment. But in the literature, it is often asserted that there is at least one confounder for these two variables. As a result, a sensitive test not looking for confounders may well mistake this edge for a 2-cycle. If the 2-cycle alpha level is set low, the direction Erk $\rightarrow$ Akt is found, as Sachs et al. predict, but as the 2-cycle alpha level is relaxed, it is eventually judged to be a 2-cycle, which could suggest an additional confounder.

With these 2-cycles marked in the graph, there are several possible cycles in the FASK model, but the set of edges in the model is nearly a superset of the set of edges in the Sachs model (with the exception of the Pka $\rightarrow$ PIP2 edge, and the reversals indicated above, for which supplemented ground truth on direction is unclear).

\section{The Mek $\rightarrow$ Erk Edge}

Sachs et al. \cite{sachs2005causal} take it as a selling point of their method (and rightly so) that it recovers the well-known Raf $\rightarrow$ Mek $\rightarrow$ Erk pathway. FASK recovers a Raf $\rightarrow$ Mek edge (along with a backward Mek $\rightarrow$ Raf edge that seems to be justified, as explained above). But it does not recover the Mek $\rightarrow$ Erk edge. Since this is a well-known connection, a comment needs to be made as to why it is not found.

\begin{figure}
\centering
\includegraphics[width=0.48\textwidth]{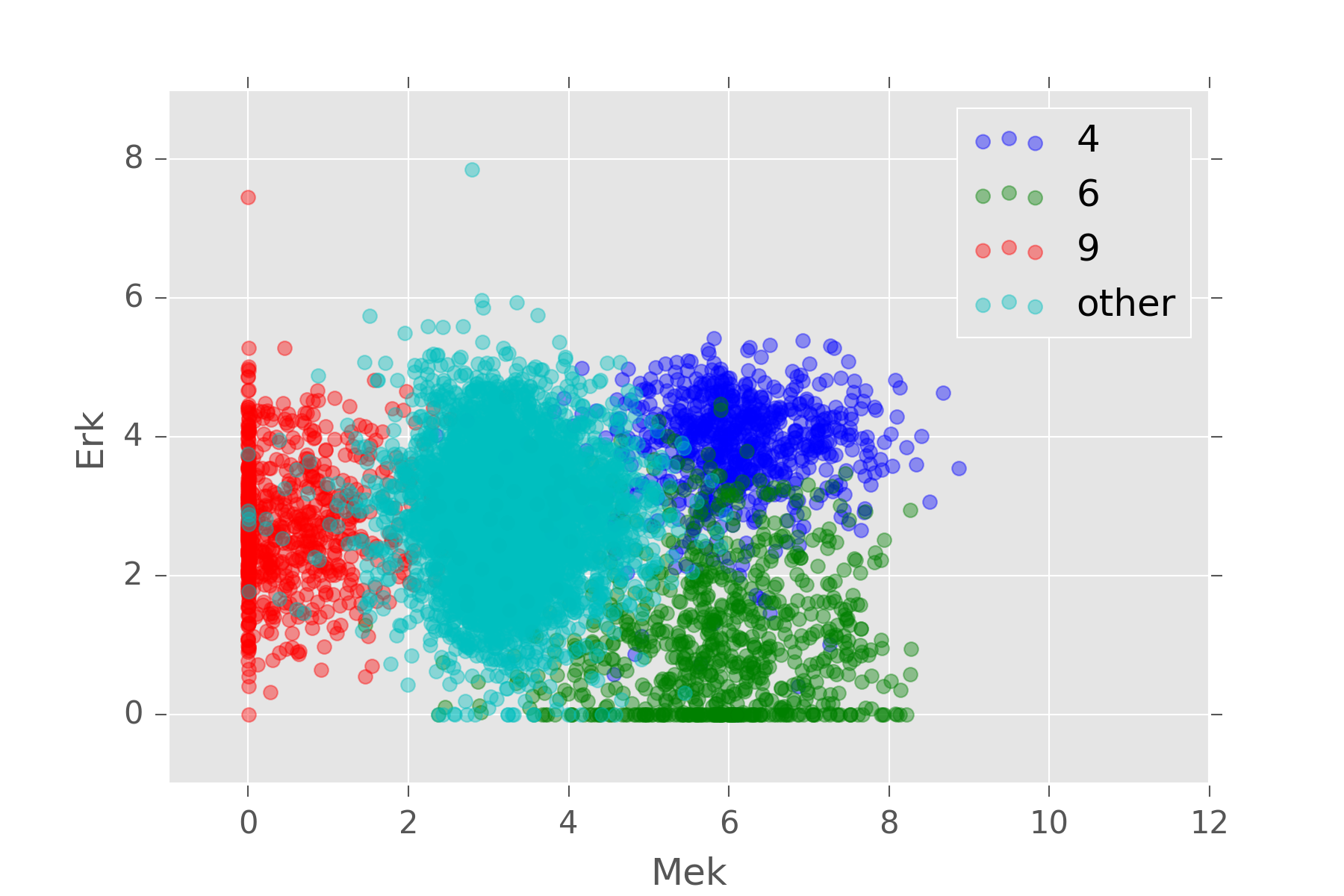}
\caption{Scatterplot of Erk versus Mek for the $\log(10 + x)$ data. See text for details.}
\label{fig12}
\end{figure}

The reason is shown in Figure \ref{fig12}. Here we show a scatter plot of Erk versus Mek, color coded by intervention. The fourth interventional context is shown in blue, the sixth interventional context in green, the nineth interventional context in red, and the rest of the interventional contexts, lumped together, in teal. What is clear, if the individual color plots are made, is that the distribution of Mek and Erk is independent for each color; an independence test shows this. From inspection, it seems that there is very little basis on which to calculate a non-zero correlation or skewness. If there is any basis at all, it is with conexts four and six. Nevertheless, overall the correlation is calculated as insignificant, and the skewness of Mek or Erk is not sufficient to warrant the addition of an edge using the heuristic skewness adjacency rule with a cutoff of 0.3. 

This appears to be a problem specific to the way FASK identifies edges, since FASK bases its judgments on correlation and (strong) skewness. One response is to use a more sensitive parameterization for FASK, and indeed if the 0.3 cutoff for the heuristic skewness adjacency rule is lowered, the edge does appear, though additional edges also appear throughout the model. (It is not clear additional edges are wrong.) Also, if this cutoff is lowered, the edge is added in the reverse direction, though other indirect paths are added from Mek to Erk.  

Another response is to consider revising FASK to take additional moments, such as kurtosis, into account. It is not clear that this would be effective; anecdotally, methods such as Two-Step \cite{sanchez2018causal} that take all moments into account also do not recover the Mek $\rightarrow$ Erk edge. This, however, requires further study.

A third response is to limit the judgment to just data from the fourth and sixth interventional contexts, where the greatest discrepancy lies. It is possible that including all of the intervention data for this example is ``washing out'' what is a already a small effect.

\section{Conclusion}

We have found an algorithm that with interventional background knowledge is able to quickly recover a graph very close to the one found in Sachs et al. \cite{sachs2005causal} using a logged preparation of their continuous data and without excluding ``outliers''\footnote{In the FASK analysis, there appear to be no outliers.}. In addition, whereas Sachs et al.'s algorithm averaged over 500 models using a heuristic search to arrive at the final model, FASK arrives at the result in one iteration, without model averaging or heuristics, even without bootstrapping. The preparation of the interventional data is very straightforward, and its incorporation into the algorithm standard. The parameters used are defaults used for other types of data (fMRI). The algorithm comes back in under a second. The result is better even than Sachs et al.'s own attempt, on one orientation, at recovering their biologically motivated extended background knowledge, which in turn is moderately to markedly better than other attempts in the literature that we have found. The best model we have found from continuous data is Miller et al. \cite{miller2012identifying}, which gives only adjacencies for the model from a nonlinear, non-Gaussian search, without orientations, so directions of causal influence are not estimated. For other published models, adjacencies and orientations are taken together noticeably worse.

An unsolved problem for FASK is the detection of latent confounders; these may show up as 2-cycles in the model for relaxed 2-cycle alphas, and it is difficult to know without further research whether these 2-cycles are genuine or represent confounding. One promising lead once again comes from the the Two Step algorithm, where confounders correspond to correlated residuals in the final model \cite{sanchez2018causal}. This method could sensibly be used to distinguish 2-cycles from confounders for FASK as well; we have not tried this tactic yet. The idea would be to estimate the FASK model using the Two-Step apparatus and check correlation of residuals, a pragmatic solution.

\bibliographystyle{abbrv}
\bibliography{bibliography}

\newpage

\section*{Appendix}
For reference, we include below models from the papers listed in the Introduction, in the same layout as earlier figures. These are the comprarisons of these models, together with the Sachs and FASK models, to the Sachs et al. ground truth.

\begin{center}
\begin{tabular}{ l | c | c | c | c }
Source & AP & AR & AHP & AHR \\
\hline
Sachs & 1.00 & 0.85 & 0.94 & 0.79 \\
FASK $10^{-5}$ & 0.84 & 0.80 & 1.00 & 0.79 \\
FASK 0.05 & 0.84 & 0.80 & 1.00 & 0.79 \\
Friedman & 0.30 & 0.30 & - & - \\
Aragam cont. &	0.50 & 0.35 & 0.43 & 0.16 \\
Aragam discr. & 0.77 & 0.50 & 0.56 & 0.26 \\
Henao & 0.90 & 0.45 & 0.67 & 0.32 \\
Miller & 0.85 & 0.85 & - & - \\
Desgranges & 0.90 & 0.45 & - & - \\
Magliacane & 1.00 & 0.30 & 0.67 & 0.21 \\
Goudet & 0.95\footnotemark & 0.90\footnotemark[7] & 0.72 & 0.68 \\
Kalainathan & 0.75 & 0.45 & 0.89 & 0.42 \\
\end{tabular}
\end{center}

These are the same models, compared to the supplemented ground truth.

\begin{center}
\begin{tabular}{ l | c | c | c | c }
Source & AP & AR & AHP & AHR \\
\hline
Sachs & 1.00 & 0.61 & 0.94 & 0.79 \\
FASK $10^{-5}$ & 0.95 & 0.64 & 1.00 & 0.79 \\
FASK 0.05 & 0.95 & 0.64 & 1.00 & 0.79 \\
Friedman & 0.53 & 0.32 & - & - \\
Aragam cont. & 0.83 & 0.36 & 0.43 & 0.16 \\
Aragam discr. & 0.92 & 0.43 & 0.56 & 0.26 \\
Henao & 1.00 & 0.36 & 0.67 & 0.32 \\
Miller & 1.00 & 0.71 & - & - \\
Desgranges & 1.00 & 0.36 & - & - \\
Magliacane & 1.00 & 0.21 & 0.67 & 0.21 \\
Goudet & 0.95\footnotemark[7] & 0.64\footnotemark[7] & 0.72 & 0.68 \\
Kalainathan & 0.92 & 0.39 & 0.89 & 0.42 \\
\end{tabular}
\end{center}

\footnotetext{Knowledge of the Sachs et al. ground truth was used to arrive at skeleton for this model.}

\begin{figure}
\centering
\begin{tikzpicture}
\tikzset{vertex/.style = {circle, draw, minimum size=2.5em,inner sep=1pt}}
\tikzset{edge/.style = {->, arrows={-> [scale=2]}}}
\tikzset{u_edge/.style = {->, arrows={- [scale=2]}}}
\node[vertex] (raf) at (-1.622, 2.524) {\small Raf};
\node[vertex] (mek) at (2.969, -0.427) {\small Mek};
\node[vertex] (plc) at (-2.729, 1.246) {\small Plc};
\node[vertex] (PIP2) at (0.845, -2.878) {\small PIP2};
\node[vertex] (PIP3) at (-0.845, -2.878) {\small PIP3};
\node[vertex] (erk) at (1.622, 2.524) {\small Erk};
\node[vertex] (akt) at (0.0, 3.0) {\small Akt};
\node[vertex] (pka) at (-2.267, -1.965) {\small Pka};
\node[vertex] (pkc) at (-2.969, -0.427) {\small Pkc};
\node[vertex] (p38) at (2.267, -1.965) {\small P38};
\node[vertex] (jnk) at (2.729, 1.246) {\small Jnk};
\draw[-] (akt) to (mek);
\draw[-] (akt) to (p38);
\draw[-] (akt) to (PIP2);
\draw[-] (jnk) to (PIP2);
\draw[-] (jnk) to (pka);
\draw[-] (mek) to (PIP2);
\draw[-] (mek) to (plc);
\draw[-] (p38) to (mek);
\draw[-] (p38) to (PIP2);
\draw[-] (p38) to (pkc);
\draw[-] (p38) to (plc);
\draw[-] (pka) to (mek);
\draw[-] (pka) to (p38);
\draw[-] (pka) to (PIP2);
\draw[-] (pka) to (plc);
\draw[-] (plc) to (PIP2);
\draw[-] (raf) to (mek);
\end{tikzpicture}
\caption{Friedman et al. \cite{friedman2008sparse}. The algorithm used here is GLASSO.}
\label{figA1}
\end{figure}
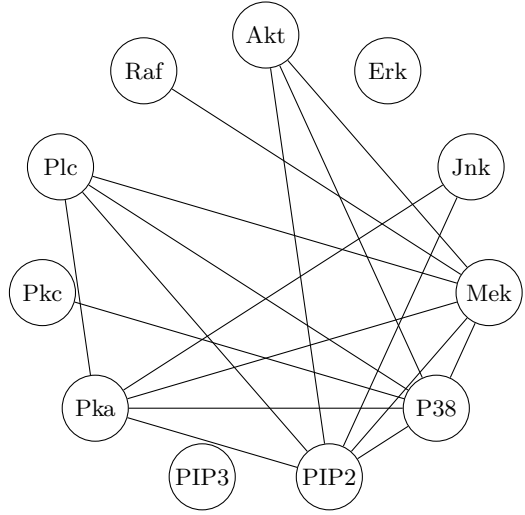

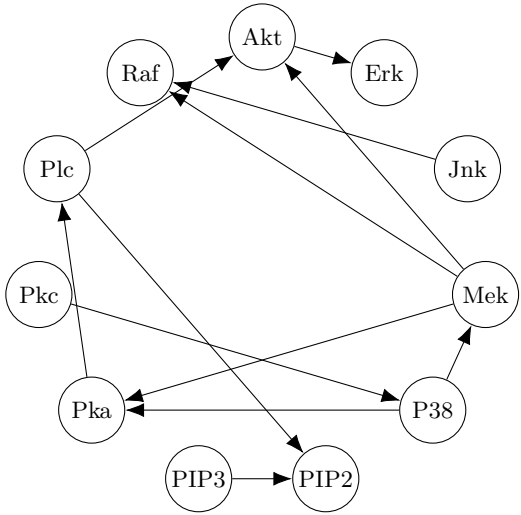
\begin{figure}
\centering
\begin{tikzpicture}
\tikzset{vertex/.style = {circle, draw, minimum size=2.5em,inner sep=1pt}}
\tikzset{edge/.style = {->, arrows={-> [scale=2]}}}
\tikzset{u_edge/.style = {->, arrows={- [scale=2]}}}
\node[vertex] (raf) at (-1.622, 2.524) {\small Raf};
\node[vertex] (mek) at (2.969, -0.427) {\small Mek};
\node[vertex] (plc) at (-2.729, 1.246) {\small Plc};
\node[vertex] (PIP2) at (0.845, -2.878) {\small PIP2};
\node[vertex] (PIP3) at (-0.845, -2.878) {\small PIP3};
\node[vertex] (erk) at (1.622, 2.524) {\small Erk};
\node[vertex] (akt) at (0.0, 3.0) {\small Akt};
\node[vertex] (pka) at (-2.267, -1.965) {\small Pka};
\node[vertex] (pkc) at (-2.969, -0.427) {\small Pkc};
\node[vertex] (p38) at (2.267, -1.965) {\small P38};
\node[vertex] (jnk) at (2.729, 1.246) {\small Jnk};
\draw[-{Latex[length=2.5mm]}] (akt) to (erk);
\draw[-{Latex[length=2.5mm]}] (jnk) to (raf);
\draw[-{Latex[length=2.5mm]}] (mek) to (akt);
\draw[-{Latex[length=2.5mm]}] (mek) to (pka);
\draw[-{Latex[length=2.5mm]}] (mek) to (raf);
\draw[-{Latex[length=2.5mm]}] (p38) to (mek);
\draw[-{Latex[length=2.5mm]}] (p38) to (pka);
\draw[-{Latex[length=2.5mm]}] (PIP3) to (PIP2);
\draw[-{Latex[length=2.5mm]}] (pka) to (plc);
\draw[-{Latex[length=2.5mm]}] (pkc) to (p38);
\draw[-{Latex[length=2.5mm]}] (plc) to (akt);
\draw[-{Latex[length=2.5mm]}] (plc) to (PIP2);
\end{tikzpicture}
\caption{Aragam et al., \cite{aragam2017learning}. Here, their algorithm is applied to the Sachs et al. continuous data.}
\label{figA2}
\end{figure}

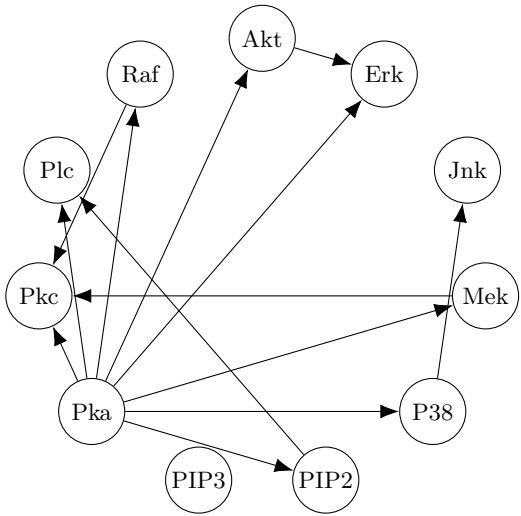
\begin{figure}
\centering
\begin{tikzpicture}
\tikzset{vertex/.style = {circle, draw, minimum size=2.5em,inner sep=1pt}}
\tikzset{edge/.style = {->, arrows={-> [scale=2]}}}
\tikzset{u_edge/.style = {->, arrows={- [scale=2]}}}
\node[vertex] (raf) at (-1.622, 2.524) {\small Raf};
\node[vertex] (mek) at (2.969, -0.427) {\small Mek};
\node[vertex] (plc) at (-2.729, 1.246) {\small Plc};
\node[vertex] (PIP2) at (0.845, -2.878) {\small PIP2};
\node[vertex] (PIP3) at (-0.845, -2.878) {\small PIP3};
\node[vertex] (erk) at (1.622, 2.524) {\small Erk};
\node[vertex] (akt) at (0.0, 3.0) {\small Akt};
\node[vertex] (pka) at (-2.267, -1.965) {\small Pka};
\node[vertex] (pkc) at (-2.969, -0.427) {\small Pkc};
\node[vertex] (p38) at (2.267, -1.965) {\small P38};
\node[vertex] (jnk) at (2.729, 1.246) {\small Jnk};
\draw[-{Latex[length=2.5mm]}] (akt) to (erk);
\draw[-{Latex[length=2.5mm]}] (mek) to (pkc);
\draw[-{Latex[length=2.5mm]}] (p38) to (jnk);
\draw[-{Latex[length=2.5mm]}] (PIP2) to (plc);
\draw[-{Latex[length=2.5mm]}] (pka) to (akt);
\draw[-{Latex[length=2.5mm]}] (pka) to (erk);
\draw[-{Latex[length=2.5mm]}] (pka) to (mek);
\draw[-{Latex[length=2.5mm]}] (pka) to (p38);
\draw[-{Latex[length=2.5mm]}] (pka) to (PIP2);
\draw[-{Latex[length=2.5mm]}] (pka) to (pkc);
\draw[-{Latex[length=2.5mm]}] (pka) to (plc);
\draw[-{Latex[length=2.5mm]}] (pka) to (raf);
\draw[-{Latex[length=2.5mm]}] (raf) to (pkc);
\end{tikzpicture}
\caption{Aragam et al., \cite{aragam2017learning}. Here, their algorithm is applied ot the Sachs et al. discrete data (the same discretization Sachs et al. use).}
\label{figA3}
\end{figure}

\begin{figure}
\centering
\begin{tikzpicture}
\tikzset{vertex/.style = {circle, draw, minimum size=2.5em,inner sep=1pt}}
\tikzset{edge/.style = {->, arrows={-> [scale=2]}}}
\tikzset{u_edge/.style = {->, arrows={- [scale=2]}}}
\node[vertex] (raf) at (-1.622, 2.524) {\small Raf};
\node[vertex] (mek) at (2.969, -0.427) {\small Mek};
\node[vertex] (plc) at (-2.729, 1.246) {\small Plc};
\node[vertex] (PIP2) at (0.845, -2.878) {\small PIP2};
\node[vertex] (PIP3) at (-0.845, -2.878) {\small PIP3};
\node[vertex] (erk) at (1.622, 2.524) {\small Erk};
\node[vertex] (akt) at (0.0, 3.0) {\small Akt};
\node[vertex] (pka) at (-2.267, -1.965) {\small Pka};
\node[vertex] (pkc) at (-2.969, -0.427) {\small Pkc};
\node[vertex] (p38) at (2.267, -1.965) {\small P38};
\node[vertex] (jnk) at (2.729, 1.246) {\small Jnk};
\draw[-{Latex[length=2.5mm]}] (erk) to (akt);
\draw[-{Latex[length=2.5mm]}] (p38) to (jnk);
\draw[-{Latex[length=2.5mm]}] (p38) to (pkc);
\draw[-{Latex[length=2.5mm]}] (PIP2) to (PIP3);
\draw[-{Latex[length=2.5mm]}] (pka) to (akt);
\draw[-{Latex[length=2.5mm]}] (pka) to (erk);
\draw[-{Latex[length=2.5mm]}] (pkc) to (jnk);
\draw[-{Latex[length=2.5mm]}] (plc) to (PIP2);
\draw[-{Latex[length=2.5mm]}] (plc) to (PIP3);
\draw[-{Latex[length=2.5mm]}] (raf) to (mek);
\end{tikzpicture}
\caption{Henao and Winther \cite{henao2011sparse}, applied to the continuous Sachs et al. data.}
\label{figA4}
\end{figure}
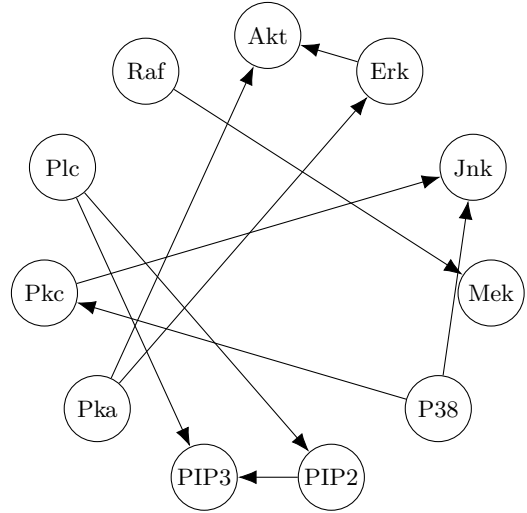

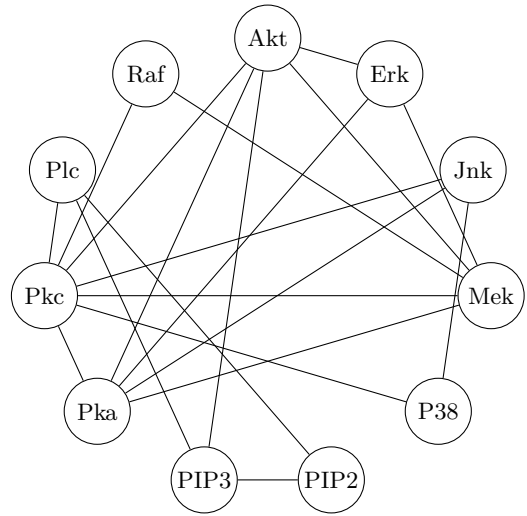
\begin{figure}
\centering
\begin{tikzpicture}
\tikzset{vertex/.style = {circle, draw, minimum size=2.5em,inner sep=1pt}}
\tikzset{edge/.style = {->, arrows={-> [scale=2]}}}
\tikzset{u_edge/.style = {->, arrows={- [scale=2]}}}
\node[vertex] (raf) at (-1.622, 2.524) {\small Raf};
\node[vertex] (mek) at (2.969, -0.427) {\small Mek};
\node[vertex] (plc) at (-2.729, 1.246) {\small Plc};
\node[vertex] (PIP2) at (0.845, -2.878) {\small PIP2};
\node[vertex] (PIP3) at (-0.845, -2.878) {\small PIP3};
\node[vertex] (erk) at (1.622, 2.524) {\small Erk};
\node[vertex] (akt) at (0.0, 3.0) {\small Akt};
\node[vertex] (pka) at (-2.267, -1.965) {\small Pka};
\node[vertex] (pkc) at (-2.969, -0.427) {\small Pkc};
\node[vertex] (p38) at (2.267, -1.965) {\small P38};
\node[vertex] (jnk) at (2.729, 1.246) {\small Jnk};
\draw[-] (erk) to (akt);
\draw[-] (erk) to (pka);
\draw[-] (jnk) to (p38);
\draw[-] (mek) to (akt);
\draw[-] (mek) to (erk);
\draw[-] (p38) to (pkc);
\draw[-] (PIP2) to (PIP3);
\draw[-] (PIP3) to (akt);
\draw[-] (pka) to (akt);
\draw[-] (pka) to (jnk);
\draw[-] (pka) to (mek);
\draw[-] (pkc) to (akt);
\draw[-] (pkc) to (jnk);
\draw[-] (pkc) to (mek);
\draw[-] (pkc) to (pka);
\draw[-] (pkc) to (plc);
\draw[-] (pkc) to (raf);
\draw[-] (plc) to (PIP2);
\draw[-] (plc) to (PIP3);
\draw[-] (raf) to (mek);
\end{tikzpicture}
\caption{Miller et al. \cite{miller2012identifying}, applied to the Sachs et al. continuous data.}
\label{figA5}
\end{figure}

\begin{figure}
\centering
\begin{tikzpicture}
\tikzset{vertex/.style = {circle, draw, minimum size=2.5em,inner sep=1pt}}
\tikzset{edge/.style = {->, arrows={-> [scale=2]}}}
\tikzset{u_edge/.style = {->, arrows={- [scale=2]}}}
\node[vertex] (raf) at (-1.622, 2.524) {\small Raf};
\node[vertex] (mek) at (2.969, -0.427) {\small Mek};
\node[vertex] (plc) at (-2.729, 1.246) {\small Plc};
\node[vertex] (PIP2) at (0.845, -2.878) {\small PIP2};
\node[vertex] (PIP3) at (-0.845, -2.878) {\small PIP3};
\node[vertex] (erk) at (1.622, 2.524) {\small Erk};
\node[vertex] (akt) at (0.0, 3.0) {\small Akt};
\node[vertex] (pka) at (-2.267, -1.965) {\small Pka};
\node[vertex] (pkc) at (-2.969, -0.427) {\small Pkc};
\node[vertex] (p38) at (2.267, -1.965) {\small P38};
\node[vertex] (jnk) at (2.729, 1.246) {\small Jnk};
\draw[-] (erk) to (akt);
\draw[-] (jnk) to (p38);
\draw[-] (jnk) to (pkc);
\draw[-] (PIP3) to (PIP2);
\draw[-] (pka) to (akt);
\draw[-] (pka) to (erk);
\draw[-] (pkc) to (p38);
\draw[-] (plc) to (PIP2);
\draw[-] (plc) to (PIP3);
\draw[-] (raf) to (mek);
\end{tikzpicture}
\caption{Desgranges \cite{desgranges2015generalization}, applied to Sachs et al. continuous data.}
\label{figA6}
\end{figure}
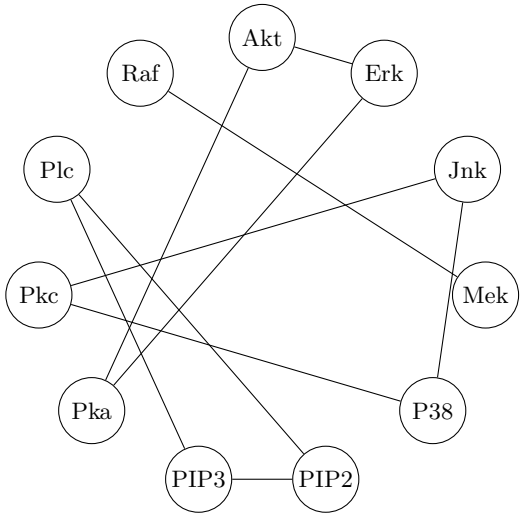

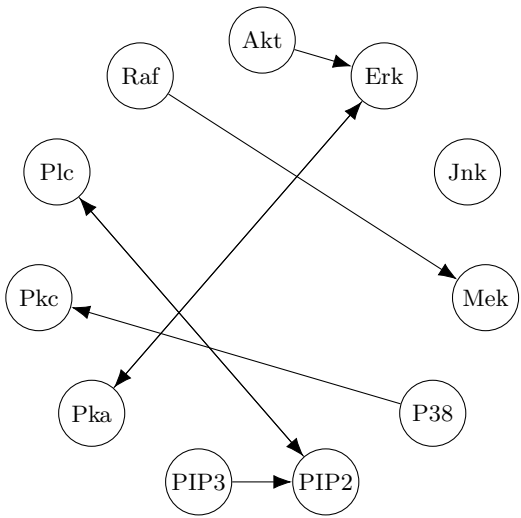
\begin{figure}
\centering
\begin{tikzpicture}
\tikzset{vertex/.style = {circle, draw, minimum size=2.5em,inner sep=1pt}}
\tikzset{edge/.style = {->, arrows={-> [scale=2]}}}
\tikzset{u_edge/.style = {->, arrows={- [scale=2]}}}
\node[vertex] (raf) at (-1.622, 2.524) {\small Raf};
\node[vertex] (mek) at (2.969, -0.427) {\small Mek};
\node[vertex] (plc) at (-2.729, 1.246) {\small Plc};
\node[vertex] (PIP2) at (0.845, -2.878) {\small PIP2};
\node[vertex] (PIP3) at (-0.845, -2.878) {\small PIP3};
\node[vertex] (erk) at (1.622, 2.524) {\small Erk};
\node[vertex] (akt) at (0.0, 3.0) {\small Akt};
\node[vertex] (pka) at (-2.267, -1.965) {\small Pka};
\node[vertex] (pkc) at (-2.969, -0.427) {\small Pkc};
\node[vertex] (p38) at (2.267, -1.965) {\small P38};
\node[vertex] (jnk) at (2.729, 1.246) {\small Jnk};
\draw[-{Latex[length=2.5mm]}] (akt) to (erk);
\draw[-{Latex[length=2.5mm]}] (erk) to (pka);
\draw[-{Latex[length=2.5mm]}] (p38) to (pkc);
\draw[-{Latex[length=2.5mm]}] (PIP2) to (plc);
\draw[-{Latex[length=2.5mm]}] (PIP3) to (PIP2);
\draw[-{Latex[length=2.5mm]}] (pka) to (erk);
\draw[-{Latex[length=2.5mm]}] (plc) to (PIP2);
\draw[-{Latex[length=2.5mm]}] (raf) to (mek);
\end{tikzpicture}
\caption{Magliacane et al. \cite{magliacane2016joint}, applied to Sachs et al. continuous data.}
\label{figA7}
\end{figure}

Figure \ref{figA1} shows the Friedman et al. model found using GLASSO in \cite{friedman2008sparse}, from continuous data. Figure \ref{figA2} shows the Aragam et al. model in \cite{aragam2017learning} from continuous data. Figure \ref{figA3} shows the model in \cite{aragam2017learning} from the same discrete data as in \cite{sachs2005causal}. Figure \ref{figA4} shows the Henao and Winther maximum likelihood model from \cite{henao2011sparse}. Figure \ref{figA5} shows the Miller et al. model from continuous data from \cite{miller2012identifying}. Figure \ref{figA6} shows the model from Desgranges' dissertation, \cite{desgranges2015generalization}. There are several models in that dissertation; the one shown here uses the KPC test with permutations. Figure \ref{figA7} shows the model from Magliacane et al. \cite{magliacane2016joint}. Figure \ref{figA7} shows the Goudet et al. \cite{goudet2017causal}, applied to Sachs et al. continuous data (only the first dataset). This method orients the edges of a provided skeleton. Here Sachs et al.'s ground truth skeleton is provided to be oriented so perhaps it is unfair to compare this graph to the other presented in this report since it start with the correct skeleton. Figure \ref{figA9} shows the Kalainathan et al. \cite{kalainathan2018sam}, applied to Sachs et al. continuous data.\footnote{There are more models of the Sachs data that have been published, undoubtedly. We did not intend to leave any out; if we did, it was an oversight. If we left yours out, please email us the reference and we will include it.}

\begin{figure}
\centering
\begin{tikzpicture}
\tikzset{vertex/.style = {circle, draw, minimum size=2.5em,inner sep=1pt}}
\tikzset{edge/.style = {->, arrows={-> [scale=2]}}}
\tikzset{u_edge/.style = {->, arrows={- [scale=2]}}}
\node[vertex] (raf) at (-1.622, 2.524) {\small Raf};
\node[vertex] (mek) at (2.969, -0.427) {\small Mek};
\node[vertex] (plc) at (-2.729, 1.246) {\small Plc};
\node[vertex] (pip2) at (0.845, -2.878) {\small PIP2};
\node[vertex] (pip3) at (-0.845, -2.878) {\small PIP3};
\node[vertex] (erk) at (1.622, 2.524) {\small Erk};
\node[vertex] (akt) at (0.0, 3.0) {\small Akt};
\node[vertex] (pka) at (-2.267, -1.965) {\small Pka};
\node[vertex] (pkc) at (-2.969, -0.427) {\small Pkc};
\node[vertex] (p38) at (2.267, -1.965) {\small P38};
\node[vertex] (jnk) at (2.729, 1.246) {\small Jnk};
\draw[-{Latex[length=2.5mm]}] (jnk) to (pka);
\draw[-{Latex[length=2.5mm]}] (mek) to (erk);
\draw[-{Latex[length=2.5mm]}] (pip2) to (pip3);
\draw[-{Latex[length=2.5mm]}] (pip2) to (pkc);
\draw[-{Latex[length=2.5mm]}] (pip2) to (plc);
\draw[-{Latex[length=2.5mm]}] (pip3) to (akt);
\draw[-{Latex[length=2.5mm]}] (pip3) to (pkc);
\draw[-{Latex[length=2.5mm]}] (pip3) to (plc);
\draw[-{Latex[length=2.5mm]}] (pka) to (akt);
\draw[-{Latex[length=2.5mm]}] (pka) to (erk);
\draw[-{Latex[length=2.5mm]}] (pka) to (mek);
\draw[-{Latex[length=2.5mm]}] (pka) to (p38);
\draw[-{Latex[length=2.5mm]}] (pkc) to (jnk);
\draw[-{Latex[length=2.5mm]}] (pkc) to (mek);
\draw[-{Latex[length=2.5mm]}] (pkc) to (p38);
\draw[-{Latex[length=2.5mm]}] (pkc) to (plc);
\draw[-{Latex[length=2.5mm]}] (pkc) to (raf);
\draw[-{Latex[length=2.5mm]}] (raf) to (mek);
\draw[-{Latex[length=2.5mm]}] (raf) to (pka);
\end{tikzpicture}
\caption{Goudet et al. \cite{goudet2017causal}, applied to Sachs et al. continuous data (only the first dataset).  This method orients the edges of a provided skeleton. Here the authors use knowledge of the Sachs et al. ground truth to come up with the provided skeleton; perhaps it is unfair to compare this model to the others presented in this report since was learned starting from a nearly correct skeleton.}
\label{figA8}
\end{figure}
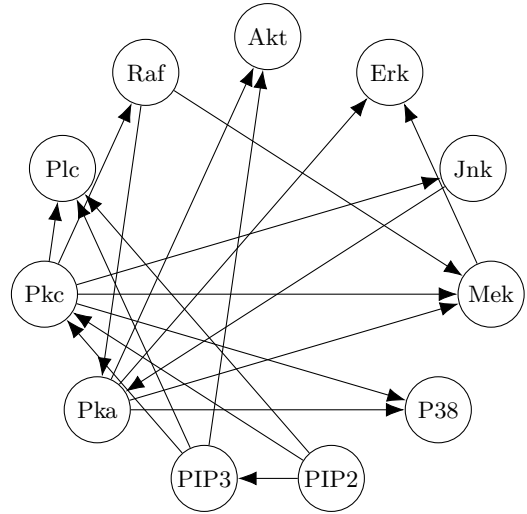

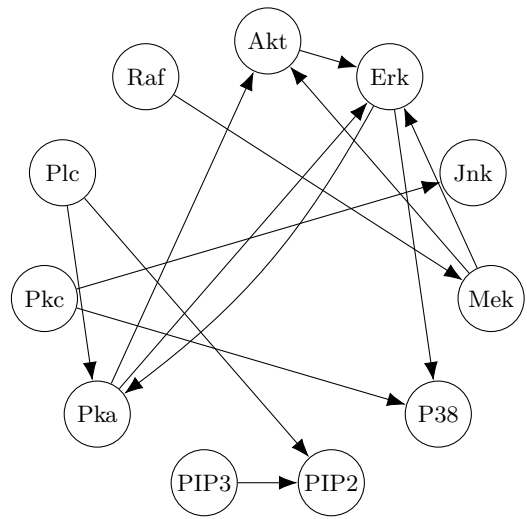
\begin{figure}
\centering
\begin{tikzpicture}
\tikzset{vertex/.style = {circle, draw, minimum size=2.5em,inner sep=1pt}}
\tikzset{edge/.style = {->, arrows={-> [scale=2]}}}
\tikzset{u_edge/.style = {->, arrows={- [scale=2]}}}
\node[vertex] (raf) at (-1.622, 2.524) {\small Raf};
\node[vertex] (mek) at (2.969, -0.427) {\small Mek};
\node[vertex] (plc) at (-2.729, 1.246) {\small Plc};
\node[vertex] (pip2) at (0.845, -2.878) {\small PIP2};
\node[vertex] (pip3) at (-0.845, -2.878) {\small PIP3};
\node[vertex] (erk) at (1.622, 2.524) {\small Erk};
\node[vertex] (akt) at (0.0, 3.0) {\small Akt};
\node[vertex] (pka) at (-2.267, -1.965) {\small Pka};
\node[vertex] (pkc) at (-2.969, -0.427) {\small Pkc};
\node[vertex] (p38) at (2.267, -1.965) {\small P38};
\node[vertex] (jnk) at (2.729, 1.246) {\small Jnk};
\draw[-{Latex[length=2.5mm]}] (akt) to (erk);
\draw[-{Latex[length=2.5mm]}, bend left=12] (erk) to (pka);
\draw[-{Latex[length=2.5mm]}] (erk) to (p38);
\draw[-{Latex[length=2.5mm]}] (pkc) to (jnk);
\draw[-{Latex[length=2.5mm]}] (pkc) to (p38);
\draw[-{Latex[length=2.5mm]}] (plc) to (pka);
\draw[-{Latex[length=2.5mm]}] (plc) to (pip2);
\draw[-{Latex[length=2.5mm]}] (pka) to (akt);
\draw[-{Latex[length=2.5mm]}] (pka) to (erk);
\draw[-{Latex[length=2.5mm]}] (mek) to (akt);
\draw[-{Latex[length=2.5mm]}] (mek) to (erk);
\draw[-{Latex[length=2.5mm]}] (pip3) to (pip2);
\draw[-{Latex[length=2.5mm]}] (raf) to (mek);
\end{tikzpicture}
\caption{Kalainathan et al. \cite{kalainathan2018sam}, applied to Sachs et al. continuous data.}
\label{figA9}
\end{figure}

\end{document}